\documentclass[12pt,nofootinbib]{revtex4}
\usepackage{amssymb,amsmath, dsfont,color}
\usepackage{graphicx}
\usepackage{epstopdf}

\newcounter{aaa}\newcounter{bbb}
\newenvironment{teor*}[2][{}]{\begin{trivlist}\refstepcounter{aaa}%
\labelsep=0pt\item[\bfseries #2. ]#1}
{\end{trivlist}}

\newcommand{\M}{r_\text{g}}
\newcommand{\A}{\mathcal{A}}
\newcommand{\T}{\mathcal{T}}
\newcommand{\h}{ h }

\newcommand{\rmd}{\mathrm{d}}

\newcommand{\obozn}{\equiv}
\newcommand{\evalat}[3]{\left.#1\right|_{#2}^{#3}}%

\newcommand{\ssy}[5]{#1,    #2 {\bf #3} (#4) #5\rlap{.}}

\begin{document}
\title{Schwarzschild-Like Wormholes as Accelerators}
\author{S. Krasnikov}
\affiliation{Central Astronomical Observatory at Pulkovo, St.Petersburg, 196140, Russia}
\email{krasnikov.xxi@gmail.com}
\begin{abstract}

 In this paper  a wormhole $W$ of an especially simple---and hence, hopefully, realistic---geometry is considered: it is static, spherically symmetric, its matter source is confined to a compact neighbourhood of the throat, and the $tt$-component (in the Schwarzschild coordinates) of its metric has a single minimum. It is shown that  such a wormhole is a  ``super  accelerator", i.~e. a pair  of free falling particles---with fixed energies at infinity---collide with the center-of-mass energy  growing unboundedly with $ |g_{tt\ \mathrm{min}}|$.The existence of such super accelerators would make it possible to observe otherwise inaccessible phenomena.

In contrast to the rotating Teo wormhole, considered by  Tsukamoto and Bambi, $W$ cannot accelerate the collision products on their way to a distant observer. On the other hand,  in contrast to the black hole colliders, $W$ does not need such acceleration  in order to make those products detectable.

\end{abstract}
\maketitle
\section{Introduction}

Suppose, in a stationary gravitational field a pair of particles with mass $m$ move on the geodesics $\gamma_{1,2}$ and then collide. The energy of the collision---let us measure it in the center-of-mass  system and  denote it as $E_{\rm c.m.}$---is not bounded by their initial  Killing energies at infinity $E_1$ and $E_2$ (to see this consider the case when both particles are initially at rest at infinity, $\varepsilon_i\obozn E_i/m=1$,  $i=1,2$: imagine two comets moving on parabolic orbits and colliding  head-on). This makes one wonder if the gain in energy can be  so large (in the vicinity of a black hole, say)  that $E_{\rm c.m.}$  gives one a chance to observe trans-Planckian effects.

The answer proved to be non-trivial.  On the one hand, it is negative  in the Schwarz\-schild space, see \cite{Pir} for the case of  $\varepsilon_{1,2}=1$. On the other hand, the energies in question do diverge in the case of the non-extreme  Kerr black holes as
 was  demonstrated by  Piran, Shaham, and Katz (PSK), who considered collisions of a special type  in those spaces and discovered \cite{Pir}  that $E_{\rm c.m.}$ of such collisions increases beyond bounds (even though $\varepsilon_{1,2}=1$) as the spin parameter $a$ tends to the mass $M$ of the black hole (i.~e., as the  black hole ``tends to the extreme one"):
\begin{equation}\label{eq:in}
\lim_{n \to\infty} a_n =M
   \quad \Rightarrow\quad   \lim_{n \to\infty} E_{\rm c.m.}(\gamma_{1n},\gamma_{2n}) =\infty
\end{equation}
(it is understood of course that the mass   $m$ does not change with $n$). Note that we speak of \emph{different} $a$'s, so it is meant that each pair $\gamma_{1n,2n}$ belongs to \emph{its own} spacetime $S_n$:
\[
a_n\neq a_{n'}   \quad \Rightarrow\quad\gamma_{1n},\gamma_{2n}\subset S_n\neq S_{n'}.
\]
As proven by Ba\~nados, Silk, and West \cite{bsw},  the divergence is ``continuous'' in the sense  that  in the limiting (extreme) spacetime  there also are geodesics $\gamma_{1n,2n}$ satisfying \eqref{eq:in} with $S_n = S$, $a_n=M$ for all $n$ (the phenomenon known as  \emph{the BSW effect}).

The PSK and BSW effects have the disadvantages that in ``reasonable astrophysical situations"
\begin{enumerate}
   \item  Kerr black holes are believed to obey ``Thorne's bound'' $a\lesssim 0.998 $ \cite{bound}, which  is  model-dependent though,  see \cite{review} for references;
   \item  it takes too long (by  the Killing time) for a particle to reach the collision point. For example \cite{review}, if the particle is  initially at rest  at a few gravitational radii from the horizon, then the corresponding  time $T$ is
 \[ \left(\frac{T}{10{\,\rm Gyr}}\right)\approx
\left(\frac{E_{\rm c.m.}}{2.5\times 10^{20}{\,\rm eV}}\right)^2   \left(\frac{M}{M_\odot}\right)
  \left(\frac{1{\,\rm GeV}}{m}\right)^2;
\]  \item \label{list:2}``\dots the photon formed from a collision just outside the
horizon can suffer a diverging redshift with decreasing
radius at a rate that exceeds the divergence of the CM
energy, and thereby results in a vanishing energy reaching infinity" \cite{NoAcc}.
 \end{enumerate}
Thus when searching for particle super accelerators one had to look at more exotic processes, such as collisions of  charged particles \cite{charged} and ``multiple scattering" \cite{GrPav,Kerr}  or more exotic spacetimes such as  five-dimensional  black holes \cite{5D}, naked singularities \cite{naked}, etc., see \cite{review} for a review. The latter list contains, in particular, the rotating wormhole \cite{B&T} considered first by Teo \cite{teo}:
\begin{multline*}
{\rm d}s^2=-N^2{\rm d}t^2+ \frac{r}{r-b}\,            {\rm d}r^2+r^2N^2\left[{\rm d}\theta^2+\sin^2\theta
({\rm d}\varphi- 2ar^{-3}\rmd t)^2\right],
\\ \text{where}\quad
N\equiv 1+{(4a\cos\theta)^2d/r}, \qquad a,b,d \text{---constants}
\end{multline*}
(from now on we use  Planck units: $G=c=\hbar=1$).
There is good reason to study the Teo wormholes: a wormhole rotating sufficiently fast, i.~e.,  having a sufficiently  large $a$, possesses an ergoregion. The Penrose effect makes such a wormhole a possible source of high energy cosmic rays.
At the same time the \emph{non-}rotating wormholes do not offer such a possibility and perhaps for this reason have not been considered so far\footnote{ The two exceptions are the Teo wormhole with $a=0$ and the Ellis wormhole. As shown in \cite{B&T} neither of them can accelerate particles unboundedly.} as  potential super colliders. The present paper  aims to fill this gap and to show by an example that the PSK effect takes place in some \emph{non}-rotating wormholes too. This may be important for the following reasons:
\begin{enumerate}\renewcommand{\theenumi}{\alph{enumi}}
  \item We do not know today whether wormholes even exist, so it is a little premature to compare their  advantages. Still, being static, spherically symmetric, and empty outside some compact region, the wormholes discussed in this paper are much simpler than Teo's, they also need much ($10^{21}$ times) less matter as source. Arguably this makes them more realistic  (following  \cite{B&T} we do  not speculate on the matter content of the wormholes, their  stability, etc.);
  \item In contrast to the situation with black holes, see item~\ref{list:2} above, high energy collisions generated by wormholes do not become invisible to a distant observer even in the absence of the Penrose acceleration. Suppose, for example, that two identical, but opposite moving, particles with energy $\varepsilon_{1}=\varepsilon_{2}$    collide at the throat  and transform into another pair of identical particles, this time with $\varepsilon_{3}=\varepsilon_{4}$. To escape the wormhole these newborn particles will have to get redshifted, indeed, but  energy conservation guarantees that they will loose exactly the same energy that was gained by the incoming ones. So, $\varepsilon_{3,4}=\varepsilon_{1,2}$: initially relativistic particles generate relativistic ejecta which, in principle, can be detected by a terrestrial observer.
\end{enumerate}

\section{Schwarzschild collider}
\subsection{The metric}
Consider a spacetime $W$ which is a Morris--Thorne  wormhole with the  metric
\begin{equation}
\label{eq:SchWh}
    \rmd s^2 =-e^{2\h(l)}\rmd t^2
+\rmd l^2
+ r^2(l) (\rmd \theta^2 +\sin^2  \theta \,\rmd \phi^2),
 \qquad l\in \mathds{R},\end{equation}
where $r(l)$ and $\h(l)$ are smooth  even functions on $ \mathds{R}^1$ monotone increasing at positive $l$ and
obeying the conditions
\begin{equation}
 \rmd r = 
\sqrt{ 1-1/x}\,\rmd l, \qquad x\obozn r/r(0)
\end{equation}
\begin{equation}\label{eq: phi}
\h(x) \evalat{}{x>3}{}
=\tfrac 12\ln( 1-1/x)
\end{equation}
(the positive constant $\M\obozn r(0)$ is a called---for obvious reasons---the radius of the wormhole). 
Thus, up to the additional condition \eqref{eq: phi} (which is added in order to make the wormhole indistinguishable by its lensing properties from the Schwarzschild black holes) the spacetime \eqref{eq:SchWh} is what was called the Schwarzschild wormhole in \cite{MTY}. It consists of the spherical layer $|l|\leq |l(r=3\M )|$  filled with matter and two empty asymptotically flat regions $r>3\M $ with  Schwarzschild metric.
%
%

\begin{figure}
\centering
\includegraphics[width= 0.7\textwidth]{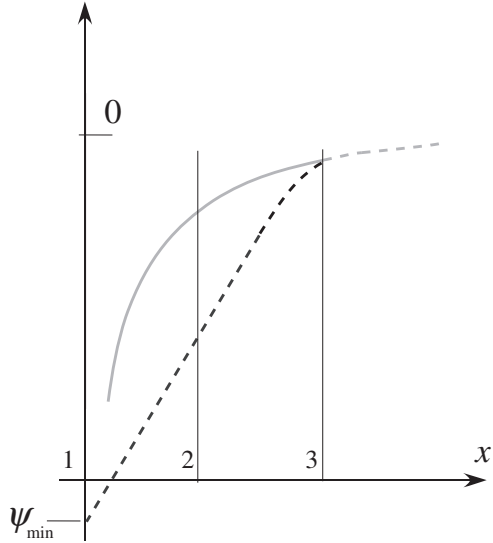}
\caption{The gray curve is the graph of $\tfrac 12\ln (1- 1/x )$. The dashed curve is $\psi(x)$.\label{fig:graph}}
\end{figure}
To specify the metric of $W$ (that is to define $h$ in the non-empty region) split the ray $x\in [1,\infty)$ into three intervals by the points $x=2$ and $x=3$, pick a positive constant $k$, and define an auxiliary  function $\psi$ by  its restrictions to those intervals:
\begin{subequations}
\begin{equation}\label{eq:x>x*}
    \psi(x)=  \tfrac 12[\ln (x-1) - \ln x]\qquad\text{at }x\in [3 , \infty);
    \end{equation}
\begin{equation}\label{eq:parabola}
    \psi(x)= -\tfrac12 (k -\tfrac{1 }{12 }) (x  - 3)^2 +\tfrac{1}{12}(x  - 3) + \tfrac 12\ln \tfrac23
\qquad \text{at }x\in [2,3 ];
\end{equation}
\begin{equation}\label{eq:x<x* -1}
     \psi (x)= k x  + c,\quad  c\obozn - 2k + \psi (2+0)
\qquad\text{at }x\in [1,2]
\end{equation}
\end{subequations}
[the  graph of $\psi (x)$ is a straight segment connected by a piece of a parabola to the Schwarzschild ``tail'' $\psi\evalat{}{x>3)}{}$, see \eqref{eq: phi} and figure~\ref{fig:graph}]. As easily seen, $\psi(x)$ is $ C^1$
\begin{align}\label{eq:ch}
   \psi (3+0)&= \tfrac 12[\ln 2 - \ln 3]= \psi (3-0), \qquad &\psi' (3+0)=&\tfrac 12[\tfrac 12 - \tfrac 13]=\psi' (3-0),\\
 \psi (2+0)&=\psi (2-0), \qquad &\psi' (2+0)=&(k -\tfrac{1 }{12 }) + \tfrac{1 }{12 }=\psi' (2-0)
\end{align}
and piecewise smooth, but its second derivative jumps at the points $x=2,3$. So, the spacetime \eqref{eq:SchWh} with $\h=\psi$ can be interpreted as describing a throat enclosed by a pair of  spherical  layers $x\in(2,3)$ with sharp boundaries\footnote{The Einstein tensor in $W $, being a combination of two first  derivatives of the metric, may have discontinuities at the spheres  $x=2,3$, on which  $\psi$ is non-smooth. However, that tensor  contains no delta-function singularities (since $\psi\in C^1$) and hence $W $ is not a ``thin-shall wormhole" \cite{viss}.}. $\psi$, however, can be arbitrarily well approximated by smooth functions,
 i.~e. for an arbitrary $\epsilon$ a smooth function $\h_\epsilon$ can be found such that
\begin{equation}\label{eq:h}
\h_\epsilon(x)\evalat{}{ [3 , \infty)}{} =\psi(x), \qquad\  |\h^{(\prime)}_\epsilon(x) - \psi^{(\prime )}(x)| < \epsilon\qquad|\h''_\epsilon(x)|\leq |\psi''(x)|,\qquad\forall x. 
\end{equation}
 We  define $W$ to be the spacetime \eqref{eq:SchWh} with $\h\obozn\h_\epsilon$,  $\epsilon$ sufficiently small, and
 $k\gg1$. The last two conditions imply, in particular, that
$c\approx -\frac52 k$ [we have substituted \eqref{eq:parabola} into \eqref{eq:x<x* -1}]
and hence 
\begin{equation}\label{eq}
    \h _\text{min}\approx  \psi(1)\approx -\tfrac32 k.
\end{equation}
\subsection{Head-on collisions at the throat}

Consider  a particle of  mass $m$ falling from the infinity $l=-\infty$ on a radial geodesic $\gamma=\Big(t(\tau),  l(\tau)\Big)\subset W$, where $\tau$ is the proper time. The spacetime is static and therefore
\begin{equation}\label{eq:eps}
    \varepsilon\obozn -g(\partial_t,\partial_\tau )=e^{2\h}\dot t \text{ \ is constant along }\gamma.
 \end{equation}
$\varepsilon$ is the initial  specific energy of the particle as measured by a Schwarzschild observer and, correspondingly, $\varepsilon\geq 1$.

Further,  by \eqref{eq:SchWh},
 \begin{equation}\label{eq:U_Sch} \dot l^2= e^{2\h }\dot t^2 -1= \varepsilon^2 e^{-2\h} -1
\end{equation}
must hold on $\gamma$. In the case at hand $\h < 0$, so on any finite interval $\dot l$ is bounded away from zero, which means that the particle never turns back\footnote{Unlike what happens sometimes in the Damour--Solodukhin wormhole \cite{foils}, which otherwise is very similar to ours.} and sooner or later reaches the throat $l=0$.

Now suppose that another, identical, particle falls in the same manner from $l=+\infty$. Then these two particles will collide at the throat.
The energy of this collision is
\[
E_{\rm c.m.}=m \sqrt{2 (1 - v^\mu_{(1)} v_{(2)\;\mu})},
\]
where $v^\mu_{(i)},$ $i=1,2$ is the velocity of the $i$-th particle at the throat, see \cite{bsw}.
Using  consecutively \eqref{eq:SchWh}, \eqref{eq:eps}, \eqref{eq:U_Sch}, and \eqref{eq} we obtain
\begin{multline}\label{eq:max}
-v^\mu_{(1)} v_{(2)\;\mu}\evalat{}{l=0}{}= e^{2\h_\text{min}} \dot t_{(1)} \dot t_{(2)} - \dot l_{(1)} \dot l_{(2)}=
\varepsilon^2 e^{-2\h_\text{min}} +\sqrt{\varepsilon^2e^{-2\h_\text{min}}-1}\sqrt{ \varepsilon^2e^{-2\h_\text{min}} -1},
\\ \approx 2\varepsilon^2 e^{-2\h _\text{min}}
 \approx 2\varepsilon^2 e^{3k}
\end{multline}
(note that the whole effect discussed in this paper originates from the sign + before the radical, which in  turn is due to the fact that $\dot l_{(1)}=- \dot l_{(2)}$ in a head-on collision).

Thus, a Schwarzschild-like wormhole  does exhibit the  PSK effect:
\begin{equation}\label{eq:Ecm}
\lim_{\h _\text{min}\to -\infty} \left(\frac{E_{\rm c.m.}}{\varepsilon m}\right)\to\infty
\end{equation}
In particular, a wormhole with
$\h _\text{min}=\ln (2m_\text{proton}/E_{\rm c.m.})\approx -44$ can  collide two initially non-relativistic ($\varepsilon_{(i)}\approx 1$) protons
 with the Planck-scale energy $E_{\rm c.m.}\approx1$.

\subsection{The technical characteristics of the collider}
In this section we roughly  estimate two parameters of our wormhole $W$. That has to be done because too large values of these parameters are obvious arguments against this type of super colliders. Define   $\A$ and $\T$ as
\begin{equation}\label{def:T}
    \T\obozn \T(1,3), \quad\text{where }\T(x_1,x_2 )\obozn
2\M \int_{x_1}^{x_2}\frac{e^{- \h} \sqrt x}{\sqrt {x-1}}\, \rmd x
\end{equation}
and
\begin{equation}\label{def:A}
\A  \obozn \A(1,\infty), \quad\text{where }
 \A(x_1,x_2)  \obozn   \M \int_{x_1}^{x_2} x^2 \max(\h,_x ^ 2(x), |\h,_{xx}(x)|) \, \frac{ \sqrt x}{\sqrt {x-1}} \,\rmd x.
\end{equation}
The chain
\begin{equation*}
\M \int_{x_1}^{x_2}\frac{e^{- \h} \sqrt x}{\sqrt {x-1}}\, \rmd x \approx
\int_{l_1}^{l_2} e^{-\h}\, \rmd l \approx\int_{l_1}^{l_2} \dot t/\dot l\, \rmd l
=t_2-t_1
\end{equation*}
shows that $\T$ is the delay (measured by a Schwarzschild observer located at $r=3\M$) between dropping a particle with $\varepsilon=1 $ into the wormhole  and receiving the same particle after it bounced off something at the throat.

The physical meaning of $\A$ is less clear.
The
scalar curvature in $W$ is
\begin{equation}\label{eq:R=1}
R=
\frac  2 {{r}^ 2 }  \Bigl(-2\, \h,_r r-
 \h,_r ^ 2 {r}^ 2 +2\,
 \h,_r ^  2\M - \h,_{rr} {r}^ 2 +
 \h,_{rr} r\M +3/2\,
 \h,_r \M \Bigr).
\end{equation}
 So, it is a sum of a few  terms each is of the order of
\begin{equation}\label{eq:est}
\M ^{-2}  \h,_x,\quad
 \M ^{-2} \h,_x ^ 2 ,\quad\text{or\ }
 \M ^{-2}\h,_{xx}
 \end{equation}
or less.
It follows from the Einstein equations and eqs.~\eqref{eq:SchWh}, \eqref{eq:est} that 
\begin{equation}\label{eq:A}
\A \gtrsim    \int_{t=0} |T_\alpha^\alpha(x)| \,\rmd^3 V,
\end{equation}
where $T_{\alpha\beta}$ is the stress-energy tensor. Thus, $\A$ can be interpreted as the ``total amount of matter'' filling the wormhole.
\begin{figure}
\centering
\includegraphics[width= 0.8\textwidth]{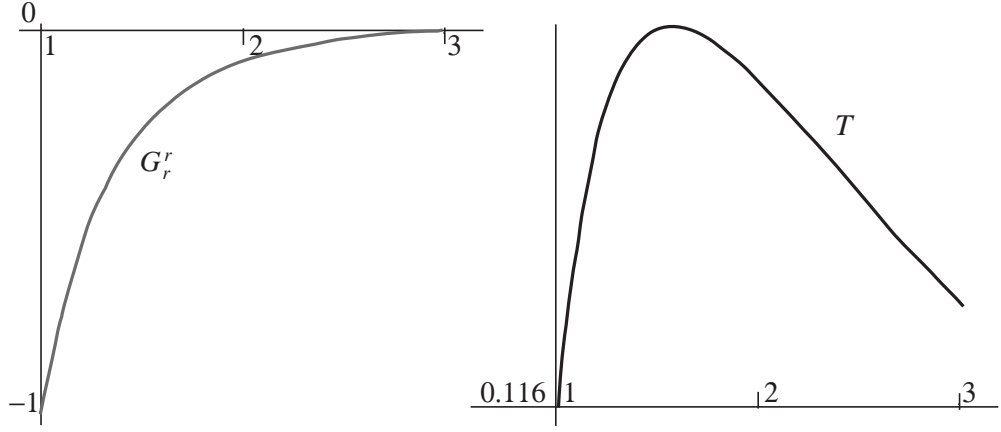}
\caption{$G^ r_{r}(x)$ and $T_\alpha^\alpha(x)$ at  $k=\frac{1}{12} $  and $m=\frac{1}{2}$ (the deformations in the vicinity of $x=2$  are neglected).
\label{fig:plot}}
\end{figure}
It is a generalization of the concept of mass to the situation   where the energy density in some non-empty regions is zero (as in our case, see eq.~(14,43) in \cite{MTW}) or even negative\footnote{We have preferred $\A$ to a more known quantifier proposed in \cite{quanti}, because the former is less coordinate dependent.}. A lot can be said against $\A$ as a characteristic of a wormhole, cf.~\cite{portal}, however, for lack of a better choice we adopt in this paper the point of view that the larger  $\A$ is, the less feasible the corresponding wormhole is.

\noindent \textbf{Remark.} There is no need in our case to discuss  separately the ``total amount of \emph{exotic} matter". Indeed, in the wormholes under consideration $G_{tt}=0$ and
\[
G_{ r r}=r^{-2}(r-2M)^{-1}[2h'r(r- M) - M],
\]
which, as shown in figure~\ref{fig:plot},  is negative at $r>2M,\ h'<0$. Thus the WEC is violated in the \emph{whole} non-vacuum part of $W$. In this sense all matter in $W$ is exotic.

\noindent \textbf{Example. The Teo wormhole.}
In this spacetime (in the case $d=0$) the scalar curvature is
\[
R = 18a^2\sin^2 \theta\, (r-b)r^{-7}
\]
and the energy of a head-on collision at the throat of two particles radially moving in the equatorial plane is
 \[
E_{\rm c.m.} \lesssim m\varepsilon
(\sqrt{ a}/b)^3 .
\]
So,
\begin{multline}\label{eq:A Teo}
\A\approx\int_{t=const} R\,\rmd^3 V\approx 10 a^2 \int_b^\infty (r-b)r^{-5}\left(1-{b\over r}
\right)^{-1/2}\,\rmd r
\\
\gtrsim
10 a^2\int _b^\infty(r-b)^{1/2}r^{-9/2}\,\rmd r = 10 a^2 b^{-3}\int _1^\infty(y-1)^{1/2}y^{-9/2}\,\rmd y
\\ \approx  a^2 b^{-3}\gtrsim  \left(\frac{E_{\rm c.m.}}{m\varepsilon}\right)^{4/3} b
\end{multline}
To get a feeling for what these estimates mean, suppose we wish to study the Planckian physics and contemplate colliding initially non-relativistic protons (which are the most available particles) accelerated by a Teo wormhole to Planck scale energies.
Then we have to take $b $ greater than the Compton wavelength of  proton (that is greater than $1/ m_\text{proton}$)  to make legitimate our classical analysis of the protons' trip through the throat. Thus,
taking into account \eqref{eq:A Teo} we need a wormhole with
\begin{equation}\label{eq:Teo's A}
\A\gtrsim  \left(\frac{m_\text{Planck }}{m_\text{proton}}\right)^{4/3} b\gtrsim
 \left(\frac{m_\text{Planck }}{m_\text{proton}}\right)^{7/3}m_\text{Planck }
\gtrsim  10^{6}M_\odot.
\end{equation}
This, as we shall see, is $10^{21}$ times more than in the case of the Schwarzschild-like wormhole.

To estimate $\T$ and $\A$ of the  wormhole $W$ we start with the segment $[2,3 ]$ (since $R=0$ at $x>3 $).
Its contribution to the total amount of matter is
\begin{multline}\A(2,3)  \lesssim  \M \int_{2}^{3} x^2 \max(\psi,_x ^ 2(x), |\psi,_{xx}(x)|) \, \frac{ \sqrt x}{\sqrt {x-1}} \,\rmd x
\\ \label{eq: A(2,3)}
\lesssim \frac 14 \M\int_2^{3} x^2k^ 2 \,  \frac{\sqrt x}{\sqrt {x-1} }\rmd x
\lesssim 10  \M k^ 2
\end{multline}
[we have used \eqref{eq:parabola} in deriving the first inequality] and its contribution to the particle's  trip time  is
\begin{equation}
\T(2,3)=2\M \int_2^{3}\frac{e^{- \h} \sqrt x}{\sqrt {x-1}}\, \rmd x\lesssim
\M \int_2^{3}  \frac{e^{-\h(2)} \sqrt x}{\sqrt {x-1}}  \,\rmd x\lesssim
2\M e^{-\h (2)}
\end{equation}
Now note that  $\psi,_x=k$ in the interval  $[1,2]$. So, again
\begin{equation}\A(1,2)  \lesssim  \M\int_1^{2} x^2k^ 2 \,  \frac{\sqrt x}{\sqrt {x-1} }\rmd x
\lesssim 10  \M k^ 2
\end{equation}
and
\begin{equation}\label{eq:T(1,2)}
\T(1,2)=\M \int_1^2\frac{e^{- \h}  \sqrt x}{ \sqrt {x-1}}\, \rmd x\lesssim
\M \int_1^2  \frac{e^{-\h _\text{min}} \sqrt x}{\sqrt {x-1}}  \,\rmd x<
10\M  e^{-\h _\text{min}}
\end{equation}
Gathering the  estimates \eqref{eq: A(2,3)}--\eqref{eq:T(1,2)} we  obtain
\begin{equation}\label{eq:sum}
 \T\lesssim 10\M  e^{-\h _\text{min}} \approx 5\M E_\text{Planck }/m_\text{proton}\approx
2 \times 10^{7}\cdot\left(\frac{\M}{3\,\text{km}}\right)\text{yr};
\end{equation}
\begin{equation}\label{eq:sum A}
   \A\lesssim 10  \M k^ 2  \approx  10^4\M
\approx 10^4\cdot M_\odot\left(\frac{\M}{3\,\text{km}}\right).
\end{equation}

\section{Conclusions}
We have considered a special class of exotic compact objects---the Schwarzschild-like wormhole with a particular profile parametrized by the radius of the throat  $\M$ and the ``depth" $\h _\text{min}$. Such wormholes have the properties of a super accelerator---two identical  particles of mass $m$ falling radially into the wormhole towards each other, collide in the throat and the energy of the collision (in the center-of-mass system) is \[E_{\rm c.m.}\approx 2\varepsilon me^{-\h _\text{min}}.
\]
So, dropping at some moment $t_0$ a relativistic proton---emitted from an accretion disk, say---in either of the mouths one makes the protons collide  at the throat with  Planck-scale energy if the wormhole has $\h _\text{min}\approx -44 $.

Collisions in kilometer-size wormholes of this type can in principle be observable.
As the  collision products escape  the wormhole they  are redshifted and come out---in tens of millions of years after $t_0$---with moderate (but still relativistic if the products are few in number) energies. Such   cosmic rays may well be registered with modern equipment.

\section*{Acknowledgements}
I am grateful to RFBR for financial support under grant No.~18-02-00461 "Rotating black holes as the sources of particles with high energy."


\begin{thebibliography}{99}
\bibitem{Pir} \ssy{T. Piran, J. Shaham, and J. Katz}{Astrophys.~J.}{196}{1975}{L107}
\bibitem{bsw}\ssy{M. Ba\~nados, J. Silk, and S. M. West}{Phys.\ Rev.\ Lett.}{103}{2009}{111102}
\bibitem{bound}\ssy{K. S. Thorne}{Astrophys.~J.}{191}{1974}{507}
\bibitem{review}\ssy{T. Harada and M. Kimura}{Classical and Quantum Gravity}{31}{2014}{243001}

\bibitem{NoAcc}\ssy{S. T. McWilliams}
{Phys.\ Rev.\ Lett.}{111}{2013}{079002}
\bibitem{charged}\ssy{O. B. Zaslavskii}{Zh.\ Eksp.\ Teor.\ Fiz.}{92}{2010}{635} (JETP Letters \textbf{92} (2010) 571).
\bibitem{GrPav}\ssy{A. A. Grib and Yu. V. Pavlov}{Grav.\ Cosmol.}{17}{2011}{42}
  \bibitem{Kerr}\ssy{S. Krasnikov and M. V. Skvortsova}{Phys.\ Rev.\ D}{97}{2018}{044019}
\bibitem{5D}\ssy{A. Abdujabbarov, N. Dadhich, B. Ahmedov and H. Eshkuvatov}{Phys.\ Rev.\ D}{88}{2013}{084036}
\bibitem{naked}\ssy{M.\ Patil and P. S. Joshi}{Class.\ Quant.\ Grav.}{28}{2011}{235012}

\bibitem{B&T}\ssy{N. Tsukamoto and C. Bambi}{Phys.\ Rev.\ D}{91}{2015}{084013; \emph{ibid.} 104040}
\bibitem{teo}\ssy{E. Teo}{Phys.\ Rev.\ D}{58}{1998}{024014}
\bibitem{MTY}\ssy{M. S. Morris, K. S. Thorne, and U.
Yurtsever}{Phys.\ Rev.\ Lett.}{61}{1988}{1446}
\bibitem{viss}M. Visser, \emph{Lorentzian wormholes --- from Einstein to
Hawking}  (AIP Press, New York, 1995).
\bibitem{foils}\ssy{T. Damour and S. N. Solodukhin}{Phys.\ Rev.\ D}{76}{2007}{024016}
\bibitem{MTW}{C. W. Misner, K. S. Thorne, and
J. A. Wheeler,  {\it Gravitation\/}  (Freeman,
San Francisco, 1973)}.
\bibitem{quanti}\ssy{M. Visser, S. Kar, N. Dadhich}{Phys.\ Rev.\ Lett.}{90}{2003}{201102}
\bibitem{portal}\ssy{S. Krasnikov}{Phys.\ Rev.\ D}{67}{2003}{104013};
S. Krasnikov, \emph{Back-in-Time and Faster-than-Light Travel in General Relativity} (Springer, Cham, 2018).
\end{thebibliography}
\end{document}